\documentclass[%
reprint,
superscriptaddress,
nofootinbib,
amsmath,amssymb,
aps,
pra,
]{revtex4-2}

\usepackage{graphicx}
\usepackage{dcolumn}
\usepackage{bm}
\usepackage{hyperref}

\usepackage{xcolor}

\begin{document}
	
	
	\title{Fully Passive Quantum Conference Key Agreement}

	\author{Jinjie Li}
	\affiliation{%
		Department of Physics, University of Hong Kong, Pokfulam Road, Hong Kong.
	}%

	\author{Wenyuan Wang}%
	\email{Corresponding author: wenyuan.wang@naist.ac.jp. Present address: Nara Institute of Science and Technology, Japan}
	\affiliation{%
		Department of Physics, University of Hong Kong, Pokfulam Road, Hong Kong.
	}%
	\affiliation{HK Institute of Quantum Science $\&$ Technology, University of Hong Kong, Pokfulam Road, Hong Kong.}

	\author{H. F. Chau}%
	\email{Corresponding author: hfchau@hku.hk}
	\affiliation{%
		Department of Physics, University of Hong Kong, Pokfulam Road, Hong Kong.
	}%
	\affiliation{HK Institute of Quantum Science $\&$ Technology, University of Hong Kong, Pokfulam Road, Hong Kong.}

	\date{\today}
	
	\begin{abstract}
		
		Quantum Conference Key Agreement (CKA) provides a secure method for multi-party communication. 
		A recently developed interference-based prepare-and-measure quantum CKA possesses the advantages of measurement-device-independence, namely, being immune to side-channels from the detector side.
		Besides, it achieves good key rate performance, especially for high-loss channels, due to the use of single photon interference. 
		Meanwhile, several fully passive QKD schemes have been proposed, which eliminate all side channels from the source modulation side. 
		We extend the fully passive idea to an interference-based CKA, which has a high level of implementation security for many-user communication.

	\end{abstract}
	
	\maketitle
	
	
	\section{INTRODUCTION}

	Quantum Key Distribution (QKD) enables secure communication between two parties by leveraging the principles of quantum mechanics, resulting in information-theoretic security.
	\cite{Bennett2014, Ekert1991}. 
	QKD has been extensively developed both theoretically and experimentally over the past few decades. 
	One major challenge in implementing QKD arises from the side channels that are inherently embedded in imperfect experimental equipment \cite{Gisin2006,Tamaki2016,Bourassa2022,Yoshino2018,Lydersen2010}.
	A novel protocol, called MDI-QKD \cite{Lo2012}, ensures immunity from all loopholes on the detector side. Later, TF-QKD \cite{Lucamarini2018, Curty2019} was proposed, which possesses the same detector side immunity and overcomes the fundamental rate-distance limit of QKD by using single-photon interference. While QKD is typically applied between two parties, efforts have been made to explore information-theoretically secure communication among many users through quantum network or quantum Conference Key Agreement (CKA) \cite{Chen2004, , Murta2020, Zhang2018, Ottaviani2019, Grasselli2018, Epping2017, Horodecki2022, Das2021, Fu2015, Li2023, Cao2021}. 
	In this paper, all CKA use quantum particles and channels.
	Notably, among the CKA protocols, several recently proposed ones are inspired by TF-QKD and are based on single-photon interference. They can achieve good key rate performance, especially under high-loss scenarios \cite{Carrara2023,MaCKA,Grasselli_2019}. 
	
	The side channels from source modulators can be eliminated by applying a recently proposed idea called fully passive QKD \cite{Wang2023a,Zapatero_2023,Marcos_2024,Lu2023, Hu2023}. 
	Such side channels may leak information to an eavesdropper or be attacked using a Trojan-horse attack \cite{Gisin2006, Yoshino2018}. 
	Passive decoy \cite{Curty2009, Curty2010} and passive encoding \cite{Curty2010} respectively remove the active modulators that generate the desired intensity or polarization of signals, replacing them with local detection and post-selection. On the other hand, fully passive QKD simultaneously post-selects both intensity and polarization, removing all modulators from the sources. This idea was first applied to BB84 \cite{Wang2023a,Zapatero_2023,Marcos_2024} and later to MDI-QKD  \cite{Li2024, Xiang_2024}. Using a different source design, two fully passive schemes for TF-QKD  \cite{Wang2023} and CV-QKD \cite{Chenyang} have also been proposed. Recently, there have also been reports of successful experimental demonstrations of fully passive QKD \cite{Hu2023, Lu2023}.
	Those works demonstrated that while enjoying the benefit of better security at the sources, not much key rate performance needs to be sacrificed. 
	
	In this work, we combine the idea of fully passive QKD with CKA, named fully passive CKA, that could possess advantages from both ideas. Our main contributions and the challenges we met of this work are as follows:
	\begin{itemize}
		\item We for the first time apply the fully passive scheme to a multi-user scenario, along with the application of a phase mis-matching strategy, results in a good sifting efficiency and reasonable performance. Fully passive protocols often suffer from heavy sifting, especially in the multi-user scenario. We show that the sifting is significantly improved using the phase mis-matching method.
		
		\item Numerical simulation of fully passive CKA is challenging due to the large amount of high-dimensional integration calculations required. However, we show that efficient computation can be achieved through the use of a high-performance integration library \cite{Hahn2005} and an optimized computer program. We also introduce a ``branch cutting'' method that significantly increases computation efficiency without sacrificing key rate performance. Details of this method can be found in the Method Section. 
		
		\item We studied several practical imperfections of the protocol, such as phase-measurement fluctuations, intensity fluctuations, and non-uniform phase distribution. Details can be found in Sections IV and V.

	\end{itemize}

	In this paper, we first briefly recapitulate both passive TF-QKD and active CKA in Sec. II, the details of the fully passive CKA protocol are explained in Sec III, including the setup preparation, how to perform the protocol, the security proof and the calculation of key rate. In Sec IV, We explicitly showed a simulated results for four-party fully passive CKA. 
	In Sec V, We provide some discussions on the protocol, as well as potential future development.

	\begin{figure}[t]
		\centering
		\includegraphics[height=10cm]{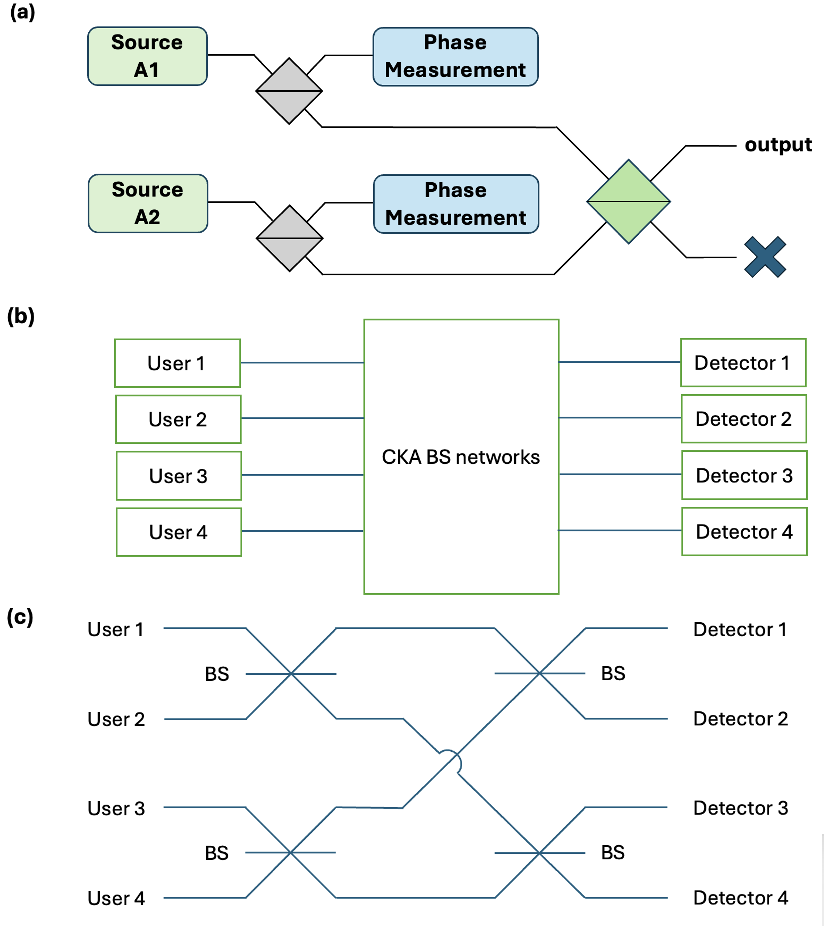}
		\caption{(a) The fully passive source setup for a single user involves the preparation of two sources by the user. The phases of these two sources are measured. The two legs are combined at a BS, and one output leg is designated as the final output of the fully passive source, based on Ref.  \cite{Wang2023}.
			(b) The schematic setup of a four-user CKA, based on Ref.  \cite{Carrara2023}.
			(c) An example of a four-user BS network, based on Ref.   }
		\label{setup}
	\end{figure}
	
	\section{Previous Works}
	\subsection{Passive TF QKD}
	Passive sources remove side-channels from the source modulator, and TF-QKD has the MDI property, which removes side-channels from the detectors. Fully passive TF-QKD has both advantages.
	
	Reference \cite{Wang2023} mainly considers a particular type of TF-QKD protocol, CAL-TF QKD \cite{Curty2019} (although the method is potentially extendable to other TF-QKD protocols too).
	In this passive TF QKD protocol, processes including encoding, decoy-state setting choices, and basis choices are all done by post-selection. The key is to use degrees of freedom coming from the randomness of passive sources.
	
	Each party's source setup is just like the one shown in Fig 1a, which will be described in the following section. Alice and Bob each make two phase measurements in their local lab. In a signal round, Alice and Bob both divide their phase domain $[0, 2\pi\}$ into $M$ slices and only post-select those rounds when their local two measurements land in the same slice and their slices are the top slice or the bottom (representing $0$ and $\pi$ phase in TF QKD). In a testing round, they post-select based on the difference of their local two measurements, representing their output intensities.

	The process works under very heavy sifting, resulting in poor key rate performance. However, in Reference  \cite{Wang2023}, a phase slice-matching strategy was introduced so that all signals could be used and contribute to the key rate. Alice and Bob send the signals to the central node, Charlie, and Charlie announces the measurement results of his two detectors, d1 and d2. The key rate is contributed by two detection patterns, for a certain slice combinations, $R(k_{A1},k_{A2},k_{B1},k_{B2}) = R = R_{1,0} + R_{0,1}$, where $R_{d_1, d_2} = p(d_1, d_2) (1 - f h(e_{X, d_1, d_2}) - h(e_{Z, d_1, d_2}) )$, where $p(d_1,d_2)$ is the probability of obtaining the click pattern $(d_1,d_2)$, $f$ is the error-correction efficiency, $h(x) = -x \log_2 x - (1-x) \log_2 (1-x)$ is the binary entropy function, and $e_{X,d_1,d_2}$ and $e_{Z,d_1,d_2}$ are the quantum bit error rate (QBER) and phase error rate, respectively.

	Due to the phase slice-matching strategy, the total key rate is the sum of all signals: \begin{equation} R = \frac{1}{M^4} \sum_{k_{A1},k_{A2},k_{B1},k_{B2}=1}^M R(k_{A1},k_{A2},k_{B1},k_{B2}), \end{equation} where $k$ represents the indices of slices into which the random phase falls, which goes from $1$ to $M$.
	
	\subsection{Conference Key Agreement}
	
	The CKA protocol allows multiple parties to conduct secure communication, and our work is based on a recently proposed CKA protocol based on single-photon interference, which is a generalization of TF-QKD. The protocol enables measurement-device-independence and can tolerate high loss \cite{Carrara2023}. 
	
	Each party prepares a WCP source and sends signals to the central node, Charlie, through a network of BSs, as shown in Fig 1. The parties send many rounds of signals, each round can be either a parameter estimation round, where parties send phase-randomized coherent lights with intensities determined by the decoy-state setting, or a key generation round, where parties send coherent states. 
	Charlie announces his detection pattern across his multiple detectors. A round is considered successful only if exactly one detector clicks, we denote this single-click event as $\Omega_j$, where $j$ indexes the specific detector that clicked. All other detection patterns are discarded. The parties then perform parameter estimation to calculate the phase error rate and QBER. 
	The key rate is calculated by:

	\begin{equation}
		R = \sum_j^M Pr(\Omega_j | PE) (1 - h(Q_Z^j)- \max _{i>0}h(Q^j_{X_0,X_i})),
	\end{equation}
	where the index $j$ represents the detector, and the index $i$ represents the party. $Q_Z^j$ is the phase error rate, and $Q_{X_0,X_i}^j$ is the QBER between party $0$ and party $i$. In this work, our key rate follows the same structural format; a detailed discussion can be found in Sec.~III~D and Appendix~B.

	\section{Method}
	
	We describe our fully passive CKA protocol, 
	which combines the single-photon-interference-based CKA protocol proposed in \cite{Carrara2023} with the fully-passive TF-QKD source and the source security analysis proposed in  \cite{Wang2023}.
	
	The fully passive CKA protocol differs from the active CKA in several key aspects:
	\begin{itemize}
		\item The application of fully passive sources introduces additional complexity to the channel model, specifically in the yield estimation from decoy state analysis, which requires certain adjustments.
		\item Unlike in the active case where parties must decide whether to perform a key generation (KG) round or a parameter estimation (PE) round, in the fully passive CKA all rounds of experiments are performed in the same manner, without the need for basis-choosing. This lack of a sifting problem is an advantage for the fully passive case, particularly when the number of users is larger.
		\item The fully passive CKA is less affected by misaligning signal preparation, which is another significant advantage.
	\end{itemize}

	\subsection{Setup}
	For a four-user fully passive CKA, each user has a fully passive source. A fully passive source, as shown in Figure 1 (a), consists of 2 laser sources that are completely random and independent of each other \cite{Wang2023}. 
	Each user measures the phases of the two pulses. The random pulses sent from the two sources interfere and the transmitted signal is the final signal sent by the user. For simplicity, throughout the paper, we assume that all users have the same intensity, $u_{max}/2$, from both laser sources. The final signal from a fully passive source can have any phase between 0 and $2\pi$ and any intensity between 0 and $u_{max}$. One could divide the whole $2\pi$ range into an even number, $M$, of sectors of the same angle, and label the top sector as sector 1, and in ascending order clockwise.

	The prepared fully passive sources then go into a CKA beam splitter (BS) network. 
	The CKA network consists of a combination of 50:50 BSs, as shown in Figure 1 (b) and (c) \cite{Carrara2023}. 
	Specifically, the BSs are arranged in $s$ layers, each layer, represented by $r = 0, 1, ...,s-1$, contains $2^{s-1}$ BSs.
	A four-user example is shown in Fig.~\ref{setup}. 
	Therefore, each layer can accept $2^s$ input signals, labeled $a^r_i$, where $i = 0,1, ... 2^s-1$ represents the position of the input site. 
	In $r$-th layer, the signals $a^r_i$ and $a^r_{i+2^r}$ interfere at the same BS. 
	
	To describe the evolution of signals within the BS network, we divide the input signals at each layer into two sets, we define the set 
	
	\begin{equation}
		F_r:=\bigcup_{k=0}^{2^{s-r-1}-1}\left\{\hat{a}_{k 2^{r+1}}, \hat{a}_{k 2^{r+1}+1} \ldots, \hat{a}_{k 2^{r+1}+2^r-1}\right\},
	\end{equation}
	and the remaining signals in the complementary set $\bar{F_r}$.
	The evolution of signals in $r$-th layer obeys
	
	\begin{equation}
		\left(\hat{a}_i^{(r)}\right)^{\dagger} \rightarrow \frac{1}{\sqrt{2}}\left[\left(\hat{a}_i^{(r+1)}\right)^{\dagger}+\left(\hat{a}_{i+2^r}^{(r+1)}\right)^{\dagger}\right]  \forall i \in F_r ,
	\end{equation}
	\begin{equation}
		\left(\hat{a}_j^{(r)}\right)^{\dagger} \rightarrow \frac{1}{\sqrt{2}}\left[\left(\hat{a}_{j-2^r}^{(r+1)}\right)^{\dagger}-\left(\hat{a}_j^{(r+1)}\right)^{\dagger}\right]  \forall j \in \bar{F}_r.
	\end{equation}
	
	Therefore, the overall evolution of the whole BS network from the input to the final  ($s-1)$-th layer satisfies
	
	\begin{equation}
		\hat{a}_i^{\dagger} \rightarrow \frac{1}{\sqrt{N_D}} \sum_{j=0}^{N_D-1}(-1)^{\vec{j} \cdot \vec{i}} \hat{d}_j^{\dagger},
	\end{equation}
	where $d_j^{\dagger}$  represents the output signals,  $\vec{j}$ is the binary representation of the detector's position, $j = 0, 1, ... ,N_D-1$, where $N_D = 2^s$ is the number of detectors. 
	The vectors $\vec{i}$ is the vector form of binary representation of the user's index, $0,1,2,...,N_U-1$, where $N_U \leq 2^s$ is the number of users \cite{Carrara2023}.

	\subsection{Protocol}
	
	1) State preparation and measurement: Each party, $A_i$, prepares the state based on the fully passive source setup and sends out the signals to the untrusted party that poccesses the BS network as shown in the Fig 1 (c). Each party locally measures the two phases, denoted $\phi_{i1}$ and $\phi_{i2}$. Charlie, the untrusted party at the detection side, publicly announces the detection results. A successful event occurs when there is exactly one click (single-click event), and all other unsuccessful events are discarded. The single-click events are labeled as $\Omega_j$, where $j$ denotes which the detector that clicks \cite{Carrara2023}. 
	
	2) Post-processing: A round of experiment is done when all parties send the fully passive signals and receive Charlie's detection announcement. 
	After a large number of rounds are performed, the first party, $A_0$, publicly announce a random binary sequence that decides if the corresponding round is a key generation (KG) round or a parameter estimation (PE) round, with probabilities $p_X$ and $1-p_X$, respectively. 
	One key advantage of our fully passive CKA protocol is that there is no basis sifting across multiple users, as each other user simply chooses the same publicly-announced basis as Alice after the transmission. 
	This is because, for our protocol, both the KG and PE round use the same physical signal, and the difference lies only in post-processing, which can be done after the transmission. 
	This is especially advantageous when the number of users is large. For $N_U$ users, the advantage of sifting efficiency over an active CKA scheme is $p_X^{N_U-1}$.

	2a) KG round: Each party, $A_i$, divides the whole $2\pi$ range into $M$ slices where $M$ must be an even number, and each of their local phase measurements should land into one of the slices indexed $k_{i1}$ and $k_{i2}$. A KG round is successful only when the slice indices obey $k_{i1} =  k_{i2} = 1$ for all $i$, or $k_{i1} =  k_{i2} = M/2+ 1$ for all $i$.
	The top slice corresponds to the plus state in X-basis, and hence corresponds to a classical bit 0, and the bottom slice corresponds to the minus state in X-basis, and hence corresponds to a classical bit 1. 
	The unsuccessful events are discarded. 
	This post-selection method discards the majority of the signals, resulting in very low sifting. However, we will show in the following sections that the discarded signals can contribute to the key rate. For now, to introduce the protocol clearly, we do not yet consider other signals.
	
	2b) PE round: Each party, $A_i$, calculates their local phase differences, $| \phi_{i1} -\phi_{i2}  |$, this is equivalent to finding their output signal intensity in the range between 0 and $u_{max}$. For example, in a 3-decoy protocol, each party divides their intensity range into 3 parts. Each party announces which part their signal falls into based on $| \phi_{i1} -\phi_{i2}  |$. In this way, all n parties essentially divide an n-dimensional square into small squares, each square, labeled $S_{ijkl}$, corresponds to a decoy state choice by all parties. They perform decoy-state analysis to estimate the yield, $Y_{n_1n_2n_3n_4}$, and subsequently calculate the phase error rate \cite{Wang2023}.
	Appendix A provides details on how to perform decoy state analysis for a two-decoy example.
	
	3) All parties perform many rounds described in step 2. 
	Their raw key is from many succesful KG rounds, each party record a classical 0 or 1 from one successful KG round. 
	All parties perform error correction and privacy amplification to obtain the final key, the same as what would be done in active CKA protocols \cite{Carrara2023}.
	
	\subsection{Security}
	
	\begin{figure}[t]
		\centering
		\includegraphics[height=5cm]{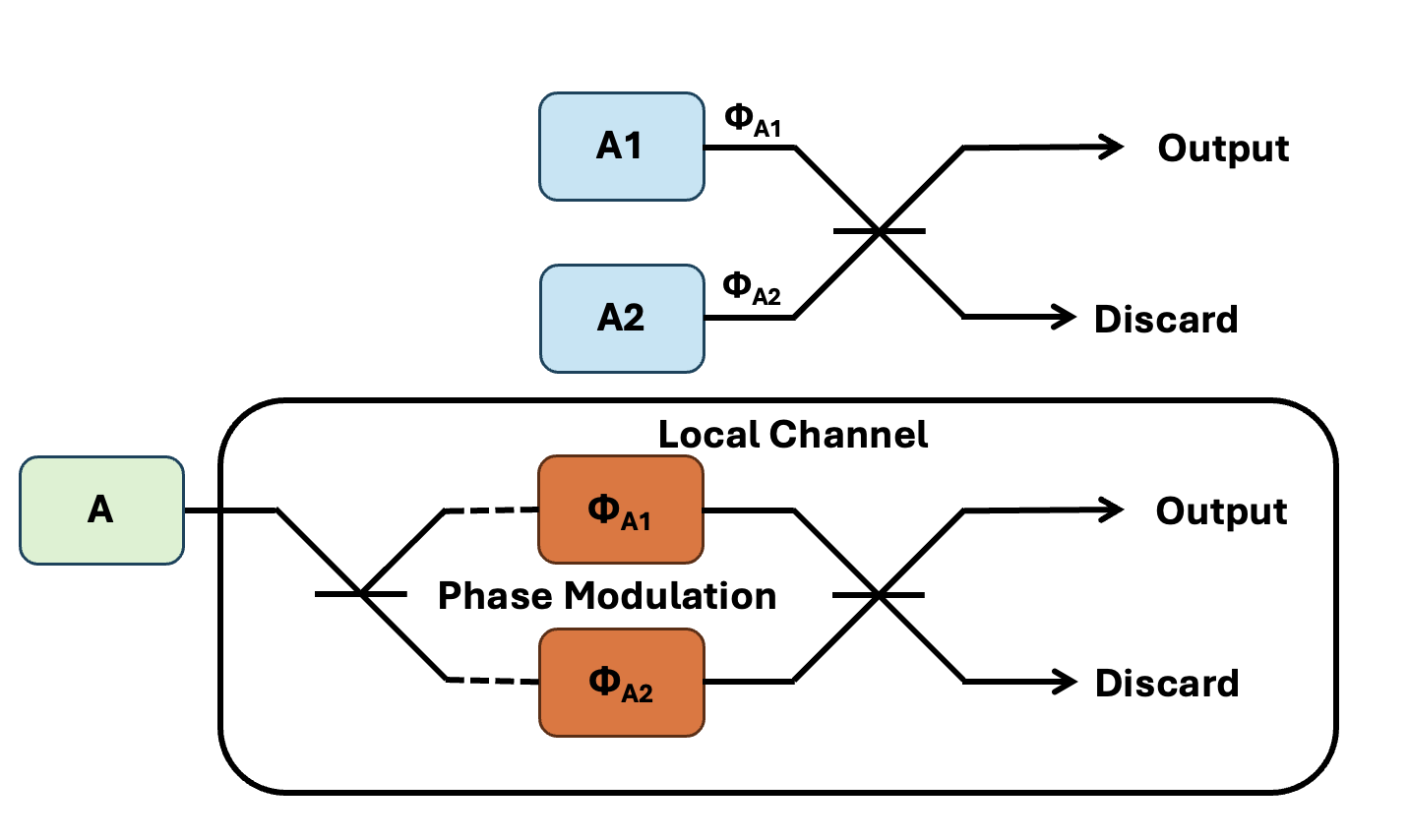}
		\caption{
			The two equivalent setups that 
			The top sub-figure is a simplified fully passive source for one user, $A$. 
			$A$ prepares two random sources, $A_1, A_2$ with phases $\phi_{A1}, \phi_{A2}$, respectively.
			Two signals join at a BS. 
			The bottom is an equivalent setup to the fully passive source. 
			Alice prepares one (perfectly prepared) source signal split evenly into two signals. 
			Then each signal goes through a random phase modulation, $\phi_{A1}, \phi_{A2}$. 
			Two signals then join at a BS. 
			The local phase modulations and the two BSs are denoted channel $\mathcal{E}_A$.
		}
		\label{security}
	\end{figure}

	The security analysis for fully passive CKA should be the same as that of active CKA, except for the imperfections in the source preparation from the fully passive sources.
	(See Appendix B for a brief summary of the secure key rate of the active CKA.)
	Instead of having a perfectly prepared signal, the phases can have any value between $[-\Delta, \Delta]$, where $\Delta = \pi/M$. This naturally causes fluctuations in the phases and intensities of the final output signals from the fully passive sources that go into the BS network. 
	
	To address this issue, we can apply the analysis from \cite{Wang2023}. One can consider an equivalent local setup that takes in two perfectly prepared signals of intensities $u_{max}/2$. These signals are split from one perfectly prepared signal, as shown in Fig. \ref{security}. 
	Each signal goes through a random phase modulation within the range $[-\Delta, \Delta]$.
	Then, the two modulated signals combine at a BS, where one of the BS output signals becomes the final output signal. Essentially, this equivalent setup replaces the two randomly generated phases inside the slice $k_{i1} = k_{i2} = 1$ with two random phase modulations, $\phi_{A1}$ and $\phi_{A2}$. 
	We name the local channel at each user's lab $\mathcal{E}_{A_i}$, which includes the two-phase modulation and the BSs.

	The most pessimistic scenario is that the entire channel is un-trusted --- we could yield the channel to the eavesdropper, Eve. 
	This would have two effects: (1) In the KG basis, QBER would be higher, and (2) in the PE basis while estimating the yields, there are additional local losses caused by the local channels $\mathcal{E}_{A_i}$ because originally, our estimation of yields from performing decoy-state analysis only represents the external channels, here, one needs to adjust them to include the effects of local channels $\mathcal{E}_{A_i}$. 
	After the adjustments, the 'corrected yields' could be used to estimate the phase error rate. 
	(The details of the corrected yields and the phase error rate calculation can be found in Appendix C. )
	
	\subsection{Key Rate} 
	
	For an N-user fully passive CKA protocol, the key rate is \cite{Carrara2023}
	\begin{widetext}
		
		\begin{equation}
			R=
			\frac{1}{M^{2N_U}}\sum_{j=0}^{N_D-1} \operatorname{Pr}\left(\Omega_j \mid \mathrm{KG}\right)\left[1-h\left(\bar{Q}_Z^j\right)-\max _{i \geq 1} h\left(Q_{X_0 X_i}^j\right)\right]
		\end{equation}
	\end{widetext}
	The sifting factor $1/M^{2N_U}$ arises from the probability that all parties' slice choices comply with the KG round procedure, as explained in Section II A, given that the total number of slices is $M$.
	$\operatorname{Pr}\left(\Omega_j \mid \mathrm{KG}\right)$ is the probability for a single-click event to happen in a KG round. 
	$\bar{Q}_Z^j$ is the phase error rate, described above, and $Q_{X_0 X_i}^j$ is the QBER for a pair of users, $A_0$ and $A_i$ \cite{Carrara2023}, 
	\begin{equation}
		Q_{X_0, X_i}^j=\operatorname{Pr}\left(X_0 \neq(-1)^{\vec{j} \cdot \vec{i}} X_i \mid \Omega_j, \mathrm{KG}\right).
	\end{equation}
	
	The sifting factor, $1/M^{2N}$, may result in a low key rate, but \cite{Wang2023} demonstrates that all other combinations can be utilized, eliminating the heavy sifting problem. 
	First, the factor can be reduced to $M^{2N-1}$ by rotating the global phase reference point freely over $0$ to $2\pi$, however, to utilize all other combinations, we need further explanation. 
	We use a two-user case to explain the stretegy and one could generalized to multi-users easily. 
	
	For two users, slice mismatching falls into four categories, none of which introduce additional loopholes. Therefore, no additional security proof is required. These categories can be mathematically represented by the four degrees-of-freedom:
	$x_1=\left(k_{A 1}+k_{A 2}\right) / 2$, $x_2=k_{A 2}-k_{A 1}$, $x_3=\left(k_{A 1}+k_{A 2}\right) / 2-\left(k_{B 1}+k_{B 2}\right) / 2$, and $x_4=\left(k_{B 2}-k_{B 1}\right)-\left(k_{A 2}-k_{A 1}\right)$. The two-user case can be generalized for N-users, where the degree of freedom is $2N$, and a similar set of $\{x_1, x_2, ... x_{2N}\}$ can represent all possible slice choices. Therefore, the key rates for all combinations can be added up  \cite{Wang2023}, for instance, in the case of a four-user CKA,
	\begin{equation}
		\begin{aligned}
			R=  \frac{1}{M^{8}}
			\sum_{k_{A 1}=1}^M \sum_{k_{A 2}=1}^M ...\sum_{k_{D2}=1}^M 
			\max \left(0, R\left(k_{A1}, k_{A2}, ... k_{D2}\right)\right)
		\end{aligned}
	\end{equation}
	where $R\left(k_{A1}, k_{A2}, ... k_{D2}\right)$ is the key rate for individual slice combinations. 
	
	While all slice choice combinations can contribute to the key rate, most of them do not provide a meaningful contribution. In particular, a combination does not contribute significantly if it satisfies one or a combination of the following conditions:
	\begin{itemize}
		\item The input phases mismatch for each party is large.
		\item The final fully passive signal phases mismatch is large for each pair of parties.
	\end{itemize}
	In the simulation results presented in the following section, we have discarded the combinations that satisfy the above conditions. This approach partially resolves the heavy computational burden involved in calculating fully passive CKA key rates. We name the method a ``branch cutting" method. (See Appendix D for details.)
	
	\section{SIMULATED RESULTS}

\begin{figure}[h]
		\centering
		\includegraphics[height=7cm]{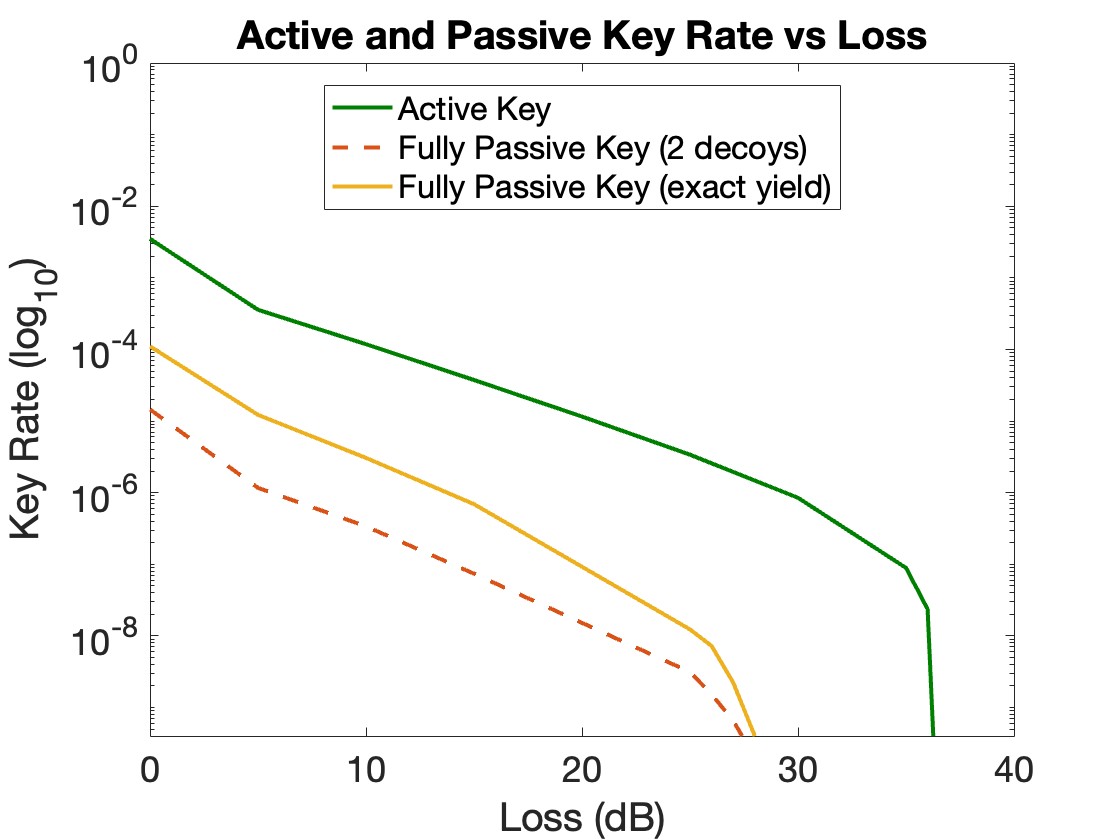}
		\caption{
			The key rate plot against channel losses between one party with the central node, for both four-user fully passive CKA and active CKA. 
			For all cases, the input intensities for all users are the same. 
			The input intensities used in the active CKA plot are optimized.
			The fully passive CKA intensities used were the same as the active ones for direct comparison. 
			The number of slices used in the fully passive case is $M = 8$, and we have applied the ``branch cutting'' method (see Sec III D).
			The phase-measurement fluctuation is also included (see Appendix E).
			The dark count used is $10^{-8}$.
		}
		\label{plot}
	\end{figure}

	Our simulation of asymptotic fully passive CKA with four users involve calculating phase error rates using both theoretical yields and linear programming to obtain upper bounds with a two-decoy setting. 
	All users have the same intensity for their fully passive sources.
	Additionally, we compare these results with those of active four-user CKA. 
	We plot key rates against the channel losses between one party and the central node, shown in Fig. \ref{plot}. 
	The results show that implementing fully passive sources reduces the key rate by around two orders of magnitude by using the theoretical yields, and a further reduction of around one order of magnitude by using two-decoy linear programming to calculate yields' upper bounds.
	In the simulation's two-decoy setting, each party divided their intensity range evenly, as described in Sec III.B. 
	The disparity in the two fully passive CKA key rate simulations could potentially be attributed to the utilization of a reduced number of decoys, which results in a looser bound on the photon number state yields in the 2-decoy case. 
	This aspect is discussed in Section V as a prospective area for future study.
	We have, for simplicity, included a phase misalignment of $0.02$~rad for one user in our simulation.
	Moreover, as we tested, applying a phase misalignment of $\pm 0.02$~rad across the users affects the key rate rather little, resulting in only an approximately $2\%$ decrease in the key rate (see Section V). 
	Meanwhile, we also included a phase-measurement fluctuation with standard deviation of 5 degrees, for details seen Appendix E. 
	The four-user fully passive CKA could reach a communication loss of about 28~dB. 
	While some communication performance is sacrificed, a much better implementation security, by applying fully passive sources, is obtained.
	In our fully passive CKA simulation, we used a  ``branch cutting'' method to alleviate the heavy computational burden,  see Appendix D for details.

	\section{DISCUSSIONS}
	In this work, we propose a fully passive CKA scheme that can simultaneously remove side channels from both the detector and source modulation sides. Furthermore, it possesses the benefit of high key rate under highly lossy channels, thanks to the application of single photon interference. Notably, this work is the first protocol proposed that possesses such a high level of implementation security for multi-party communication. 
	
	Our simulation shows that the key rate compared to an active CKA is about two to three orders of magnitude lower due to the nature of heavy sifting in fully passive schemes. (three to four orders of magnitude lower while using linear programs to perform decoy-state analysis. )
	However, better implementation security at the source side is achieved, and therefore a reasonable amount of key rate sacrifice is acceptable.
	Just like fully passive TF-QKD \cite{Wang2023}, our fully passive CKA scheme has some important additional advantages over the active case, including:
	\begin{itemize}
		\item There is no need to choose a basis, as there is no sifting required. This is because the procedure for performing a KG round and a PE round is identical.
		\item Our protocol possesses an inherent resistance to large phase misalignments, similar to fully passive TF-QKD \cite{Wang2023}. This arises from the symmetry in the \textit{local} phase slice reference point for each user. Because the protocol sums the key rate contributions from all phase positions from $0$ to $2\pi$, a misalignment that shifts a user away from their optimal "zero-position" phase slice simply maps them onto another valid phase slice on the circumference. 
		
		We highlight two important nuances regarding this symmetric behavior. First, because the phase space is divided into discrete segments (a slice size of $\pi/4$ when $M=8$), a substantial channel misalignment of $\pi/4 + \delta$ simplifies to a minor misalignment of just $\delta$. Conversely, small misalignments of $\delta$ occurring strictly within a slice boundary cannot be fully cancelled and will introduce a slight increase in the QBER. Second, to reduce computational overhead in our simulations, we utilized a branch-cutting technique (detailed in Appendix D) that restricts the evaluated phase slices to those near the zero position. Although this approach reliably models realistic, small-scale errors (such as a test with misalignments of $0$, $+0.02$, $-0.02$, and $-0.02$~rad, caused only an around $2\%$ drop in the key rate at both 5~dB and 10~dB, applying $[x=y=3]$ branch cut), it computationally breaks the full rotational symmetry, meaning very large misalignments would not cancel in this specific simulation setup. 
		If we would like to handle large misalignments, in principle, one can turn off the branch-cutting setup and include all the phase slices in the calculation.

	\end{itemize}
	These advantages are particularly useful in the case of CKA, since a larger number of users would normally induce heavier sifting and more difficulties in aligning.
	
	Several implementation imperfections may need some brief comments here: For the phase distribution, gain-switched lasers were proven to be feasible in experiments for both high-speed MDI-QKD \cite{Woodward2021} and fully passive BB84 QKD \cite{Hu2023, Lu2023}. Intensity distributions (which represent phase distributions in passive protocols) and random tests have shown that gain-switched lasers follow the uniform and independent assumptions well. Even if non-uniform phase distributions arise, as long as we can characterize the phase distribution of the source, even if it is non-uniform, and if we can accurately measure the phase of each pulse, we can perform an additional round of post-selection to selectively discard signals with different probabilities depending on the measured phase. This strategy, proposed in \cite{Wang2023a}, aims to "shape" it into a uniform distribution. However, accurately measuring the phase presents another challenge, which we will discuss below. 
	
	On the other hand, if phase correlations are present, a recent work \cite{Marcomini2024} has studied the potential higher-order correlations in gain-switched lasers and proposed a security analysis to address this, which is, in principle, also applicable to our scheme. 
	The intensity distributions have demonstrated close adherence to theoretical predictions, suggesting a near-uniform phase distribution \cite{Lu2023, Hu2023}. Moreover, randomness tests have revealed minimal self-correlation among intensities, indicating a similar lack of correlation among phases; the robust performance of gain-switched lasers for such randomness generation is well-established in both QKD and quantum random number generation \cite{Comandar2016, Abellan2014}.

	In Appendix E, we investigate how phase-measurement fluctuations may impact performance. Our analysis shows that when the phase measurement follows a Gaussian distribution with a standard deviation of 5 degrees, the key rate decreases by approximately 10\%.

	Furthermore, it is important to note the impact of practical receiver-side and channel imperfections, such as detector-efficiency mismatches and out-of-mode misalignments (e.g., polarization, time, and spatial misalignments). Because our fully passive CKA protocol employs an MDI-type architecture with a completely untrusted central node, receiver-side imperfections cannot open side-channels and therefore do not compromise the security proof. They enter solely through the channel model, affecting only the protocol's performance. To evaluate this degradation in a realistic, asymmetric setting, we simulated the four-user protocol with detector efficiency mismatches of 1, 0.94, 0.97, and 0.92, alongside slightly varying dark count rates of $1 \times 10^{-8}$, $1.2 \times 10^{-8}$, $0.9 \times 10^{-8}$, and $1.1 \times 10^{-8}$. We also incorporated out-of-mode misalignments of 0, 0.05, -0.03, and 0.04 rad, respectively. Using the $[x=3, y=3]$ branch cut setting, the key rate decreased by only ${\sim}1\%$ at both 5~dB and 10~dB channel losses. This demonstrates that the performance degradation in a realistic setting with receiver-side imperfections is highly negligible.

	This work is an important proof-of-concept that a fully passive scheme can be applied to a multi-user communication scheme without losing much key rate performance. However, there is room for further improvement in key rate performance when higher performance computers are available. These improvements include the optimization of the slice number $M$, and input intensity, as well as performing a full calculation of all slice choices without using the ``branch cutting'' method.

	An important next step, with higher computational power, is to calculate the performance for more users. Specifically, the scaling law between the number of users and the sifting caused by the application of fully passive sources is planned for future study. Additionally, an experimental realization of the protocol is a subsequent step, and it will be the first fully passive many-user experiment.
	Incorporating a three-decoy state analysis may result in tighter yield upper bounds and increase key rate performance, therefore, it remains to be implemented in the future work.

	For finite-size analysis, we cite the results from \cite{Curty} (which proposed a finite-analysis for active CAL TF-QKD) and \cite{Wang2023} (which extended it to passive CAL TF-QKD). Notably, \cite{Wang2023} demonstrated that practical performance can be achieved for passive TF-QKD with reasonable data sizes (e.g., $N = 10^{13}$). In its analysis, the Z-basis decoy-state estimation remains minimally affected as it only needs to be performed once and is shared across all key generation sets, while smaller X-basis signal data (due to post-selecting phase slices) is a more unique challenge for passive TF-QKD. However, \cite{Wang2023} showed that, using concentration inequalities \cite{Curty}, one can still obtain an acceptable bound on the phase error rate. This approach is, in principle, applicable to passive CKA, too (which has a structure similar to CAL TF-QKD, and it also allows the sharing of decoy statistics across all key generation patterns), though a rigorous study will be a subject for future work.

	\begin{acknowledgments}
		We thank H.-K. Lo.  for insightful comments.  
		H.F. Chau is supported by the RGC Grant No. 17303323 of the HKSAR Government. 
		W. Wang acknowledges support from the University of Hong Kong Seed Fund for Basic Research for New Staff, the Hong Kong RGC General Research Fund (No. 17312922), and the NSFC Young Scientist Fund (No. 12204391).
	\end{acknowledgments}

	\appendix

	\section{Decoy State Analysis - Two-Decoy}
	\begin{widetext}
		Decoy-state analysis is utilized to determine the upper bounds of the yields, denoted as $Y^{Z,j}_{m_A m_B m_C m_D}$, by applying linear programs to the set of equations, in the four-user case \cite{Wang2023, Carrara2023}:
		\begin{equation}
			G^j_{S_{ijkl}}=\sum_{m_A, \ldots, m_D=0}^{\infty} Y_{m_A, \ldots, m_D}^j 
			\left\langle P\left(m_A, m_B, m_C, m_D\right)\right\rangle_{S_{ijkl}}
		\end{equation}
		where $G^j_{S_{ijkl}}$ is a direct observable from the channel model, representing the conditional probability of a single-click event in a PE round, given the decoy state setting $S_{ijkl}$, $G^j_{S_{ijkl}} = \text{Pr}(\Omega_j|S_{ijkl})$ \cite{Carrara2023}.
		
		The average photon number distribution is given by
		\begin{equation}
			\begin{gathered}
				\left\langle P^Z\left(m_A, \dots, m_D\right)\right\rangle_{S_{ijkl}}=
				\frac{1}{P_{S_{i j jk }}} \int_{S_{i j kl}} 
				P_{\text {Poisson }}\left(\mu_A, m_A\right) \dots P_{\text {Poisson }}\left(\mu_D, m_D\right)
				P_{\text {int }}\left(\mu_A, \dots , \mu_D\right) d \mu_A \dots d \mu_D,
			\end{gathered}
		\end{equation}
		where $	P_{\text {Poisson }}$ is the Poissonian distribution, and 
		\begin{equation}
			P_{S_{i j jk }} = \int_{S_{i j kl}}  P_{\text {int }}(\mu_A,\dots,\mu_D)d \mu_A \dots d \mu_D,
		\end{equation}
		\begin{equation}
			P_{\text {int }}\left(\mu_A, \dots ,\mu_D\right)=\frac{1}{\pi^4 \sqrt{\mu_A\left(\mu_{\max }-\mu_A\right)} \sqrt{\mu_B\left(\mu_{\max }-\mu_B\right)} \sqrt{\mu_C\left(\mu_{\max }-\mu_C\right)} \sqrt{\mu_D\left(\mu_{\max }-\mu_D\right)}}
		\end{equation}
		is the intensity probability distribution \cite{Wang2023}.
		The bounds obtained from the linear program can then be applied to Equation B1 for local loss correction and subsequently to calculate the phase error rate.
		
	\end{widetext}

	\section{Active CKA Key Rate\cite{Carrara2023}}
	
	The lower bound of the asymptotic key rate of active CKA with one-way reconciliation, under collective attacks can be written as 
	
	\begin{align}
		r   \geq &
		\sum_{j=0}^{M-1} \operatorname{Pr}\left(\Omega_j \mid \mathrm{KG}\right) \nonumber \\& \quad
		\times  
		\left[H\left(X_0 \mid E\right)_{\Omega_j}-\max _{i \geq 1} H\left(X_0 \mid X_i\right)_{\Omega_j}\right]
	\end{align}
	where each post-selected succesful events,  $\Omega_j$, corresponds to a seperate key, and we add them up to form the final key. Here, $M$ is the number of detectors. 
	The term $H\left(X_0 \mid E\right)$ is the entropy of the outcome of a KG round of the first party $A_0$ given the information held by the eavesdropper.
	The term $H\left(X_0 \mid X_i\right)$ is the entropy of the outcome of a KG round of the first party $A_0$ given another party $A_i$'s outcome.

	\section{Security}

	The imperfect preparation from the two inputs of the fully passive sources can be regarded as two-phase modulation  $[-\Delta, \Delta]$, as shown in \ref{security}. 
	As mentioned in the main text, we pessimistically yield the entire channel $\mathcal{E}_{A_i}$, which includes the phase modulations and the two BSs, to Eve.
	Under this assumption, one obvious effect is the KG basis QBER will be larger. 
	Another effect is that we need to consider the effects of the local channels on PE, specifically, the estimation of yields. 
	
	Taking a four-user fully passive CKA as an example, estimates of yields, denoted as $Y^{Z,j}_{m_A m_B m_C m_D}$, can be obtained by performing decoy state analysis. The output photon numbers are represented by $m_A$, $m_B$, $m_C$, and $m_D$ (see Appendix D), while $j$ indicates the single-photon event $\Omega_j$.
	Here, we explicitly use the superscript $Z$ to indicate that it is in the $Z$ basis for clarity. However, it will be omitted in subsequent equations.
	While considering the local channel losses, one corrects the yield in the following way, 
	\begin{widetext}
		\begin{align}
			Y_{n_A n_Bn_C n_D}^{j}= 
			& \int_{-\Delta_\phi}^{\Delta_\phi} \int_{-\Delta_\phi}^{\Delta_\phi} ...\int_{-\Delta_\phi}^{\Delta_\phi} 
			\sum_{m_A=0}^{n_A} \sum_{m_B=0}^{n_B}\sum_{m_C=0}^{n_C}\sum_{m_D=0}^{n_D}
			P_{\phi_{A 1}, \phi_{A 2}}\left(m_A \mid n_A\right) 
			P_{\phi_{B 1}, \phi_{B 2}}\left(m_B \mid n_B\right) \nonumber 
			\\
			&  \quad
			P_{\phi_{C1}, \phi_{C 2}}\left(m_C\mid n_C\right) 
			P_{\phi_{D1}, \phi_{D2}}\left(m_D \mid n_D\right) 
			\quad  Y_{m_A m_Bm_Cm_D}^{j} d \phi_{A 1} d \phi_{A 2} ... d \phi_{D2} 
		\end{align}where the local photon loss from  $m_A, m_B , m_C , m_D$ to  $n_A, n_B , n_C , n_D$ that is represented by a conditional probability $P_{\phi_{A 1}, \phi_{A 2}}\left(m_A \mid n_A\right)$, using party A as an instance \cite{Wang2023}.
		\begin{equation}
			P_{\phi_{A 1}, \phi_{A 2}}\left(m_A \mid n_A\right)
			= \left| \frac{1}{2!} \sqrt{\frac{n_{A}!}{m_{A}!\left(n_A-m_A\right)!}}  
			\left[1+e^{i \pi / 2+\left(\phi_{A 2}-\phi_{A 1}\right)}\right]^{m_A}
			\left[1+e^{i \pi / 2+\left(\phi_{A 2}-\phi_{A 1}\right)}\right]^{\left(n_A-m_A\right)} \right|^2,
		\end{equation}
		the yield then can be used to calculate the phase error rate just as active CKA \cite{Carrara2023}:
		
		\begin{equation}
			\begin{aligned}
				\bar{Q}_Z^j=\frac{1}{\operatorname{Pr}\left(\Omega_j \mid \mathrm{KG}\right)} \sum_{v \in \mathcal{V}} 
				\left(\sum_{n_0+\cdots+n_{N-1} \leq \bar{n}} \prod_{i=0}^{N-1} c_{i, n_i}^{\left(v_i\right)} \sqrt{\bar{Y}_{n_0, \ldots, n_{N-1}}^j}+\Delta_{v, \bar{n}}\right)^2,
			\end{aligned}
		\end{equation}
	\end{widetext}
	where $\operatorname{Pr}\left(\Omega_j \mid \mathrm{KG}\right)$ is the conditional probability that a single-click event happens given that the round is a KG round, $N$ is the number of users, and $\bar{n}$ is the cutoff photon number used to perform decoy state analysis, and
	\begin{equation}
		\mathcal{V} =\left\{v \in\left\{0,2^N-1\right\}:|\vec{v}| \quad \bmod 2=0\right\},
	\end{equation}
	\begin{equation}
		\begin{aligned}
			c_{i, n}^{(l)} = \begin{cases}
				e^{-\alpha_i^2 / 2} \frac{\alpha_i^n}{\sqrt{n!}} & \text{if } n+l \text{ is even}, \\
				0 & \text{if } n+l \text{ is odd},
			\end{cases} \\
			\label{}
		\end{aligned}
	\end{equation}
	\begin{equation}
		\Delta_{v, \bar{n}}  =\sum_{n_0+\cdots+n_{N-1} \geq \bar{n}+2} \prod_{i=0}^{N-1} c_{i, n_i}^{\left(v_i\right)} .
	\end{equation}

	\section{Branch Cutting Method}
	In fully passive CKA, in order to speed up the calculation, we introduce this method to not calculate some slice-choice combinations because those do not contribute meaningful key rate, or have zero key rate.
	These discarded combination satisfy at least one of the following criteria: 
	
	\begin{itemize}
		\item The input phases mismatch for each party is large.
		Specifically, for party $i$, let $k_{i1}$ and $k_{i2}$ be her two slice choices, we require $| k_{i1} -k_{i2}| \leq x $ for all $i$.
		\item The final fully passive signal phases mismatch is large for each pair of parties.
		Specifically, for a pair of parties $i$ and $j$, we demand 
		\begin{equation}
			\frac{k_{i1}+k_{i2}}{2} - 		\frac{k_{j1}+k_{j2}}{2} \leq y,
		\end{equation}
		for all pairs. 
	\end{itemize}
	Here, $x$ and $y$ must be a positive integer that is less than the total slice number $M$. 
	In this paper, $M = 8 $ and $x=y=3$ was used in the simulation shown. 
	
	Here, we present in Table 1 some tests we carried out to observe the tradeoff between computational time and key rate by choosing different values of $x$ and $y$:
	\begin{table}[h]
		\centering
		\begin{tabular}{|c|c|c|}
			\hline
			Branch cutting choice & Key rate & Computational time \\
			\hline
			$x = y = 2$ & 1 & 1 \\
			$x = 2, y = 3$ & 4 & 5 \\
			$x = 3, y = 2$ & 6 & 8 \\
			$x = 2, y = 4$ & 10 & 23 \\
			$x = y = 3$ & 15 & 35 \\
			\hline
		\end{tabular}
		\caption{Tradeoff between computational time and key rate for different branch cutting choices.} \label{table:table1}
	\end{table}

	\section{Phase Measurement Fluctuations}
	We also conducted a simulation to investigate how phase-measurement fluctuations affect key rate performance. In this simulation, we incorporated a Gaussian measurement distribution with a standard deviation of $5^\circ$. By applying a significance level of three standard deviations on each side, we observed an approximate 10\% reduction in key rate performance.
	
\begin{figure}[h]
		\centering
		\includegraphics[height=11cm]{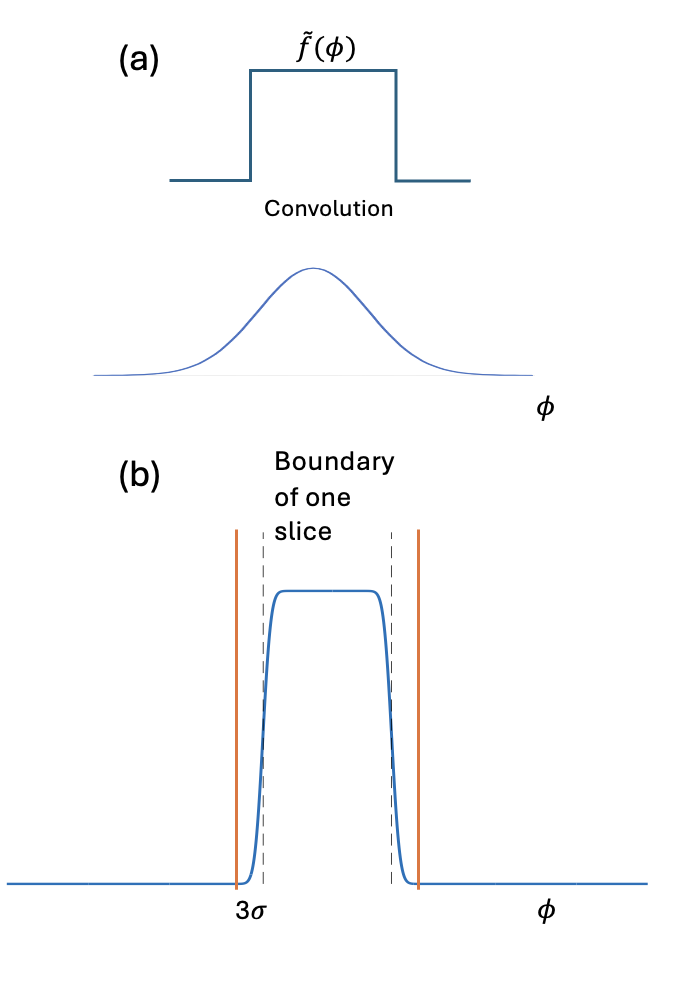}
		\caption{
			\textbf{(a)} The truncated distribution $\tilde{f}(\phi)$ and a  Gaussian distribution. 
			\textbf{(b)} The resulting distribution after convolution. Dashed lines denote the original slice boundaries, while orange lines indicate the integration limits extended by $3\sigma$.
		}
		\label{fluc}
	\end{figure}

	During the simulation, we convolved the phase probability distribution with the Gaussian measurement distribution for each slice. 
	Specifically, we first truncated the overall phase distribution $f(\phi)$, originally defined over $[0, 2\pi]$, to create a modified distribution $\tilde{f}(\phi)$ [see Fig.~\ref{fluc}(a)].
	The truncated phase distribution $\tilde{f}(\phi)$ is explicitly defined as:
	$$
	\tilde{f}(\phi) = 
	\begin{cases}
		f(\phi), & \text{if } \phi \in [\phi_{\text{slice}}^-,\, \phi_{\text{slice}}^+], \\
		0, & \text{otherwise},
	\end{cases}
	$$
	where $[\phi_{\text{slice}}^-,\, \phi_{\text{slice}}^+]$ denotes the angular range of the slice of interest, and $f(\phi)$ is the original phase probability distribution.
	We then convolved $\tilde{f}(\phi)$ with the Gaussian distribution $G(\phi)$:

	\begin{align*}
		(\tilde{f} * G)(\phi) = \int_{-\infty}^{\infty} \tilde{f}(\phi') G(\phi - \phi') \, d\phi'.
	\end{align*}
	
	The resulting distribution was integrated using extended limits, widened by three standard deviations ($3\sigma$) on each side compared to the original integration range [see Fig.~\ref{fluc}(b)].

	Including phase-measurement fluctuations and applying finite integration limits led to two effects: (1) a small loss of signal in the distribution tails, and (2) a slight increase in quantum bit error rate (QBER). These factors collectively resulted in a minor reduction in the achievable key rate.

	\newpage

	\bibliography{ref1.bib}
	
\end{document}